\documentclass[12pt]{iopart}

\usepackage[english]{babel}
\usepackage{graphicx}

\begin{document}

\title[Parameter estimation for compact binary inspirals into LIGO data]{Parameter estimation for signals from compact binary inspirals injected into LIGO data}

\author{Marc van der Sluys$^{1}$, Ilya Mandel$^{1}$,  Vivien Raymond$^{1}$,
  Vicky Kalogera$^{1}$, Christian R\"{o}ver$^{2}$, Nelson Christensen$^{3}$}
\address{$^1$ Physics \& Astronomy, Northwestern University, Evanston IL, USA}
\address{$^2$ Max-Planck-Institut f\"ur Gravitationsphysik, Hannover, Germany}
\address{$^3$ Physics \& Astronomy, Carleton College, Northfield MN, USA}
\ead{sluys@northwestern.edu}

\begin{abstract}
  During the fifth science run of the Laser Interferometer Gravitational-wave Observatory 
  (LIGO), signals modelling the gravitational waves emitted by coalescing non-spinning compact-object 
  binaries were injected into the LIGO data stream.  We analysed the data segments into 
  which such injections were made using a Bayesian approach, implemented as a Markov-chain 
  Monte-Carlo technique in our code \textsc{SPINspiral}.  This 
  technique enables us to determine the physical parameters of such a binary inspiral, 
  including masses and spin, following a possible detection trigger.  
  For the first time, we publish the results of a realistic parameter-estimation analysis 
  of waveforms embedded in real detector noise.  
  We used both spinning and non-spinning waveform templates for the data analysis and
  demonstrate that the intrinsic source parameters can be estimated with an accuracy 
  of better than 1--3\% in the chirp mass and 0.02--0.05 (8--20\%) in the symmetric mass ratio if 
  non-spinning waveforms are used. We also find a bias between the
  injected and recovered parameters, and attribute it to the difference
  in the post-Newtonian orders of the waveforms used for injection and analysis.
\end{abstract}

\submitto{Classical and Quantum Gravity}

\section{Introduction}
\label{sec:intro}

The inspiral of stellar-mass compact binaries via gravitational-wave emission
is among the most promising gravitational-wave sources for ground-based laser 
interferometers, such as LIGO~\cite{1992Sci...256..325A,statusofligo2006}
and Virgo~\cite{statusofvirgo2007}. If such a binary contains a black hole, 
it is expected to be spinning moderately~\cite{2008ApJ...682..474B}.  A 
spinning black hole causes the binary orbit to precess, introducing phase and 
amplitude modulations in the gravitational-wave signal. 
Accounting for these effects improves the detection efficiency and also improves 
the accuracy of the parameter estimation on the signal. 
The accuracy with which the binary parameters can be determined is of significant 
astrophysical interest; the effect of spin on this accuracy is described
in an initial parameter study \cite{2008ApJ...688L..61V}.

We have developed a code called \textsc{SPINspiral} which implements a Markov-chain Monte-Carlo (MCMC) 
technique~\cite{gilks_etal_1996} to compute the posterior probability-density 
functions (PDFs), or simply \textit{posteriors}, of the source parameters
of binary inspirals with spinning components.  
\textsc{SPINspiral} is a modification of an 
earlier parameter-estimation code for analysis on binaries without 
spin~\cite{2007PhRvD..75f2004R, RoeverThesis2007}.  In addition to including 
post-Newtonian gravitational waveforms with a single spinning 
object~\cite{1994PhRvD..49.6274A}, 
we have also implemented a number of improvements designed to make the 
parameter-space exploration more efficient~\cite{2008CQGra..25r4011V,2009_methods_paper}.

During the fifth LIGO science run (S5)~\cite{2006CQGra..23S..51S}, modelled 
gravitational-wave signals from compact binary coalescences with non-spinning 
members were injected into LIGO data, in order to test the detection 
pipeline~\cite{2008CQGra..25r4006G,2009arXiv0905.0020L}.  These hardware 
injections were made for a variety of masses and signal-to-noise ratios (SNRs). 
The data sets thus obtained provide the most realistic approximations to actual 
observed gravitational-wave signals and form an ideal testbed for our parameter-estimation
code.  In particular, the noise realisation is that of the actual interferometers,
as opposed to the Gaussian noise we have used in previous studies, and the injected 
waveforms are of a higher (2.0) post-Newtonian (pN) order.

In this paper, we present the results of the parameter-estimation analysis
with \textsc{SPINspiral} of three such hardware injections; the signals of
binary inspirals of black-hole--neutron-star (BH-NS), NS-NS and
BH-BH binary inspirals.  In \sref{sec:implement} we briefly
describe our MCMC code.  \Sref{sec:analysis} contains the results of
our analyses, where we show that we typically recover the chirp mass to a few 
percent accuracy, while in some cases we find a bias between the injected and
recovered values of mass and spin parameters.  We conclude these proceedings 
with a discussion of these results in \sref{sec:conclusions}.

\section{Implementation of MCMC technique}
\label{sec:implement}

In this section, we briefly describe the main features of our MCMC code \textsc{SPINspiral} for binary 
inspirals with spin, including the waveform used, the likelihood computation and the way we
choose our priors and perform jump proposals.  More details can be found in
\cite{2008CQGra..25r4011V} and in a forthcoming paper \cite{2009_methods_paper}.

\subsection{Waveform}
\label{sec:waveform}

We use a waveform
that takes into account post-Newtonian (pN) expansions up to the 1.5-pN
order in phase and is restricted to the Newtonian order in amplitude.
The waveform includes the simple-precession prescription~\cite{1994PhRvD..49.6274A}.
This choice of waveform template allows us to investigate the first-order
effects of spin (spin-orbit coupling), as long as only one binary
member has spin.  In comparison to higher-order pN, double-spin waveforms, 
where the spin-spin interaction is taken into account, this waveform can be computed analytically
and has a parameter space with lower dimensionality, so that the computational cost per iteration is 
lower and the number of iterations needed for convergence is smaller.

The waveform for a binary inspiral with one spinning object is described by
12 parameters: the chirp mass $\mathcal{M} \equiv \frac{(M_1 M_2)^{3/5}}{(M_1+M_2)^{1/5}}$,
symmetric mass ratio $\eta \equiv \frac{M_1 M_2}{(M_1+M_2)^2}$, 
the constant, dimensionless spin magnitude $a_\mathrm{spin,1} \equiv S_1/M_1^2$,
the constant angle between spin and orbital angular momentum $\theta_\mathrm{spin,1}$,
the luminosity distance $d_\mathrm{L}$ and sky position $\alpha$, $\delta$,
the time, orbital phase and precession phase at coalescence $t_\mathrm{c}$, 
$\varphi_\mathrm{c}$, $\varphi_\mathrm{spin,1}$, and two angles that define the
direction of the total angular momentum $\vec{J}_0$ of the binary: 
$\iota_\mathrm{J_0}$ and $\psi_\mathrm{J_0}$.
Each waveform template is computed in the time domain, and then windowed and Fourier 
transformed.  The calculation of the likelihood \eref{eq:likelihood}, 
which measures how well a model waveform matches the data, is carried out in the frequency domain.

\subsection{Computation of the likelihood}
\label{sec:likelihood}

We follow a Bayesian approach to infer the posterior probability-density
functions (PDFs) of the twelve parameters that describe our waveform.
We can compute the likelihood for a model waveform $\tilde{m}(\vec{\lambda},f)$
with parameters $\vec{\lambda}$ and data set $\tilde{d}(f)$ as measured by 
a detector $i$ using
\begin{equation}
  L_i(d|\vec{\lambda})  \propto \exp \left( -2 \int_0^\infty  \frac{\left|\tilde{d}(f) 
    - \tilde{m}(\vec{\lambda},f) \right|^2}{S_\mathrm{n}(f)}\, df  \right).
  \label{eq:likelihood}
\end{equation}
The tildes indicate that both $d$ and $m$ are expressed in the frequency domain,
and $S_\mathrm{n}(f)$ is the one-sided noise power-spectral density.
Assuming that the noise of different interferometers is independent, the 
expression for the posterior PDF given the data from a coherent network of $N$ interferometers 
generalises to
\begin{equation}
  p(\vec{\lambda}|d) \propto p(\vec{\lambda}) \, \prod_{i=1}^{N} L_i(d|\vec{\lambda}),
  \label{eq:posterior}
\end{equation}
where $p(\vec{\lambda})$ is the \textit{prior} distribution of the parameters.
Further details on Bayesian inference and its application to a coherent network of 
detectors can be found in \cite{2007PhRvD..75f2004R}.

\subsection{Prior distributions}
\label{sec:priors}

We use prior distributions that are uniform in $\log(d_\mathrm{L})$, 
$\cos(\theta_\mathrm{spin,1})$, $\sin(\delta)$, $\cos(\iota_\mathrm{J_0})$, 
and in the linear scales of the remaining parameters.
\textsc{SPINspiral} is designed to work as a follow-up analysis tool, 
to be used \textit{after} a gravitational-wave trigger has been detected and identified
as a binary-inspiral candidate.  Hence, we can rely on some prior information from the detection
trigger, in particular for the chirp mass $\mathcal{M}_\mathrm{tr}$ and time of coalescence
$t_\mathrm{c,tr}$ of the trigger.
We start our Markov chains from a chirp-mass value chosen randomly from a 
Gaussian distribution with a width of 0.25\,$M_\odot$\footnote{This value
is somewhat arbitrary, and relatively large. If chains can converge from such
distant initial values, they are guaranteed to converge when started from the 
value returned by the detection pipeline.} around $\mathcal{M}_\mathrm{tr}$,
with a value for $t_\mathrm{c}$ from a Gaussian that is 0.1\,s wide around 
$t_\mathrm{c,tr}$, and randomly from the whole prior distribution in all other
parameters.

The prior ranges used in this study for the 12 parameters are between 
$\mathcal{M} \!\in\! [0.5\,\mathcal{M}_\mathrm{tr}, 2.0\,\mathcal{M}_\mathrm{tr}]$;
$\eta \!\in\! [0.03,0.25]$;
$t_\mathrm{c} \!\in\! [t_\mathrm{c,tr}\!-\!50\,\mathrm{ms}, t_\mathrm{c,tr}\!+\!50\,\mathrm{ms}]$;
$d_\mathrm{L} \!\in\! [10^{-3}\,\mathrm{Mpc}, 100\,\mathrm{Mpc}]$;
$a_\mathrm{spin,1} \!\in\! [0,1]$; 
$\cos(\theta_\mathrm{spin,1}) \!\in\! [-1,1]$;
$\varphi_\mathrm{spin,1} \!\in\! [0,2\pi]$;
$\alpha \!\in\! [0,2\pi]$; 
$\sin(\delta) \!\in\! [-1,1]$; 
$\cos(\iota_\mathrm{J_0}) \!\in\! [-1,1]$;
$\psi_\mathrm{J_0} \!\in\! [0,\pi]$; and
$\varphi_\mathrm{c} \!\in\! [0,2\pi]$.
We use these broad, flat priors to keep our study general. However,
when additional information becomes available (for instance the time and
sky location of a gamma-ray burst during an externally-triggered search), 
the priors can be restricted accordingly.

\subsection{Jump proposals}

Bayes' theorem allows us to combine the model's prior
distribution and likelihood and derive the PDF containing the information 
about the parameters given the data. 
We use stochastic Monte-Carlo integration for this purpose, implemented in a Markov-chain Monte-Carlo
(MCMC) scheme (\textit{e.g.}\ \cite{gilks_etal_1996}). The Metropolis 
algorithm used here requires the specification of a proposal distribution, 
which defines how (and how efficiently) the parameter space is explored. 
We use an adaptive scheme to regulate the proposed jump 
size (\textit{e.g.}\ \cite{atchade_rosenthal_2005}) with a target acceptance
rate of about 25\%.
  
Typically we propose approximately 10\% of our MCMC jumps for each parameter
individually.  The other jump proposals are done in all 12 parameters at once.
To make these all-parameter updates more efficient, we use the covariance
matrix that the code computes from previous iterations to guide the jump
proposals.  More details of our jump proposals are described in section~2.5 of
\cite{2008CQGra..25r4011V} and in \cite{2009_methods_paper}.

In order to both explore all of the wide prior ranges of our parameter space 
and at the same time sample the much smaller regions of interest in detail, we
use a technique that is known as \textit{parallel tempering}.  We run several
(typically $\sim 5$) Markov chains in parallel, each with a different
`temperature' assigned.  
While hot chains can explore a larger part of the parameter space, only data
from the coolest chain, which samples the regions of interest in detail,
is used to generate the posterior PDFs.  
Our implementation of parallel tempering is described in more detail in section~2.6
of \cite{2008CQGra..25r4011V} and in \cite{2009_methods_paper}.

\section{MCMC analysis of hardware injections}
\label{sec:analysis}

During the fifth LIGO science run (S5), which ended in October 2007, modelled 
gravitational-wave signals from compact binary coalescences (CBCs) were injected into 
LIGO data to test the CBC detection pipeline.  The signals
were introduced in the detector by physically moving the mirrors of the interferometer,
according to precomputed simulated 2.0-pN, inspiral waveforms.
Such \textit{hardware injections} were done for a variety of binary-component masses 
and for different signal-to-noise ratios (SNRs).  

Most of these hardware injections had non-spinning components, and were done simultaneously 
for the LIGO 4-km interferometers H1 (Hanford) and L1 (Livingston), and 2-km interferometer 
H2.  However, the signal injected into each detector was that of an optimally oriented (`face-on')
source located exactly overhead that detector.  Hence, the injections were not coherent, 
\textit{i.e.}, there is no predefined choice of sky location, orientation and distance 
for the source that is consistent in all interferometers.

Since the SNR in the H2 detector is typically about half of that in the other 
two LIGO interferometers, adding the H2 data to that of H1 and L1 in our coherent 
analysis raises the network SNR by only about 6\%, while it increases the 
computational cost by 50\%.  The CPU time scales linearly with the number of 
detectors included in the analysis.  We therefore 
ran our parameter estimation with the data from the H1 and L1 only.\footnote{Strictly 
speaking, using data from both H1 and H2 would invalidate Eq.~\ref{eq:posterior}, 
since the noise realisations of the two detectors are not independent.}

In order to accurately sample the parameter space in our MCMC 
runs, we typically need to accumulate a few million iterations.  
A run with two detectors and the waveform described in \sref{sec:waveform} 
takes about a week to compute $2\times10^6$ 
iterations on a single 2.8\,GHz CPU, when five parallel-tempering chains
are used.

\subsection{A black-hole--neutron-star injection}
\label{sec:bh-ns}

In this section, we present our analysis of the hardware injection 
of the signal from a non-spinning inspiral of a 10.0\,$M_\odot$ BH and 
a 1.4\,$M_\odot$ NS at a distance of 50\,Mpc.  We used 7.0\,s of 
data from the H1 and L1 interferometers for the parameter estimation.  
We ran the analysis both allowing the BH spin to be determined (spMCMC) and 
assuming that no spin was present (nsMCMC). 

\begin{figure}
  \resizebox{\textwidth}{!}{
    \includegraphics[angle=0]{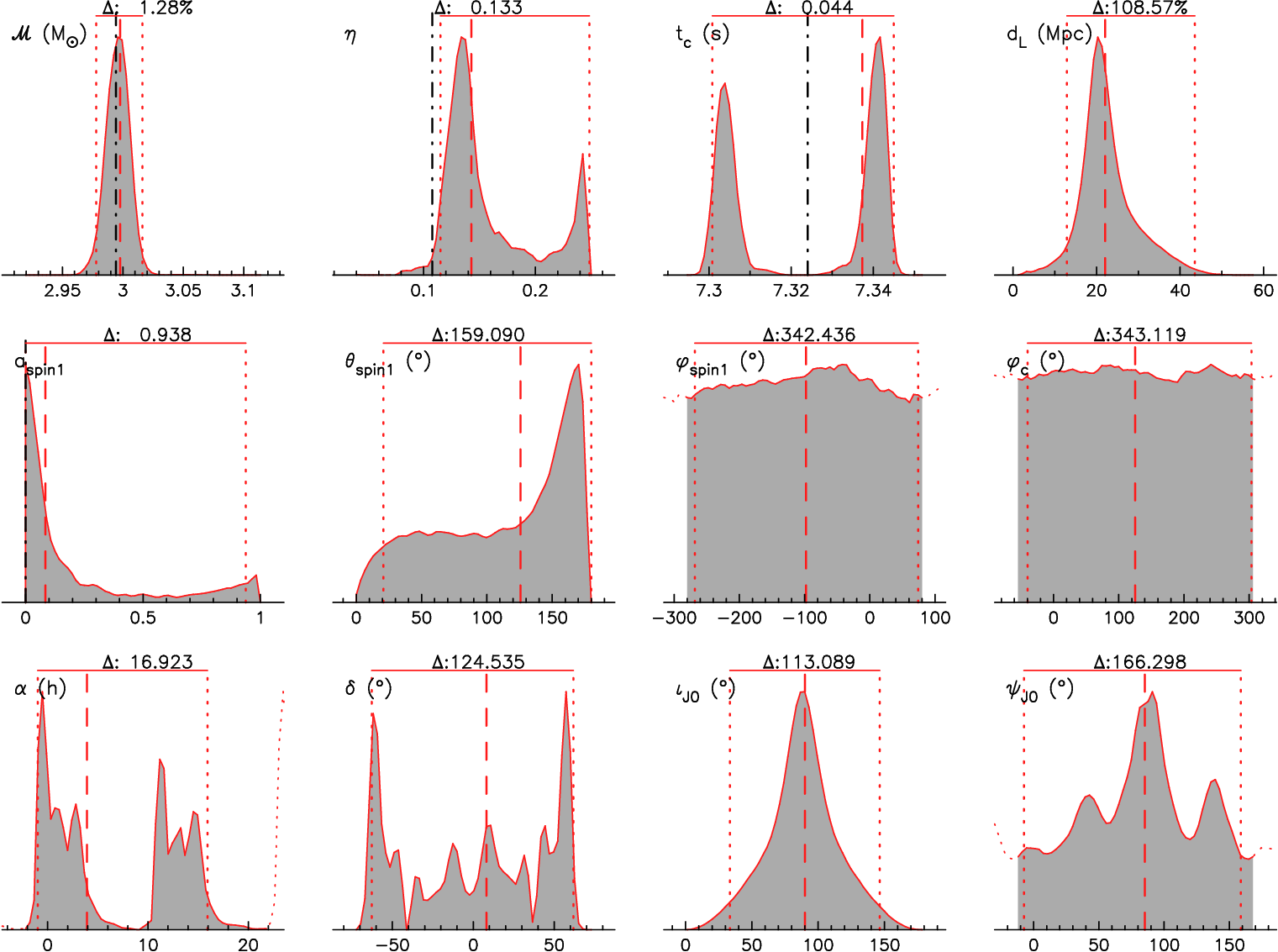}
  }
  \caption{
    Results of the spMCMC analysis of the BH-NS hardware injection. Each panel shows
    the posterior PDF for one of the 12 parameters.  The dotted lines indicate
    the 2-$\sigma$ probability range, the value of which is shown in the top of
    the panel as $\Delta$ (the relative values are with respect to the centre of
    the range).  The dashed line shows the median of the distribution, the dash-dotted 
    line indicates the parameter value of the injection where available for a
    non-coherent injection.
  }
  \label{fig:bh-ns_pdf1d}
\end{figure}

\Fref{fig:bh-ns_pdf1d} shows
the marginalised PDFs from the spMCMC analysis.  Each of the panels in
the figure contains the posterior PDF for one of the twelve parameters.
The railing of the PDF for $a_\mathrm{spin1}$ against the lower limit 
of the prior range is consistent with the non-spinning nature of the injection.
The 2-$\sigma$ probability ranges shown in the figure are defined as the 
smallest range that contains 95.45\% of the PDF.  

\begin{figure}
  \resizebox{\textwidth}{!}{
    \includegraphics[angle=0]{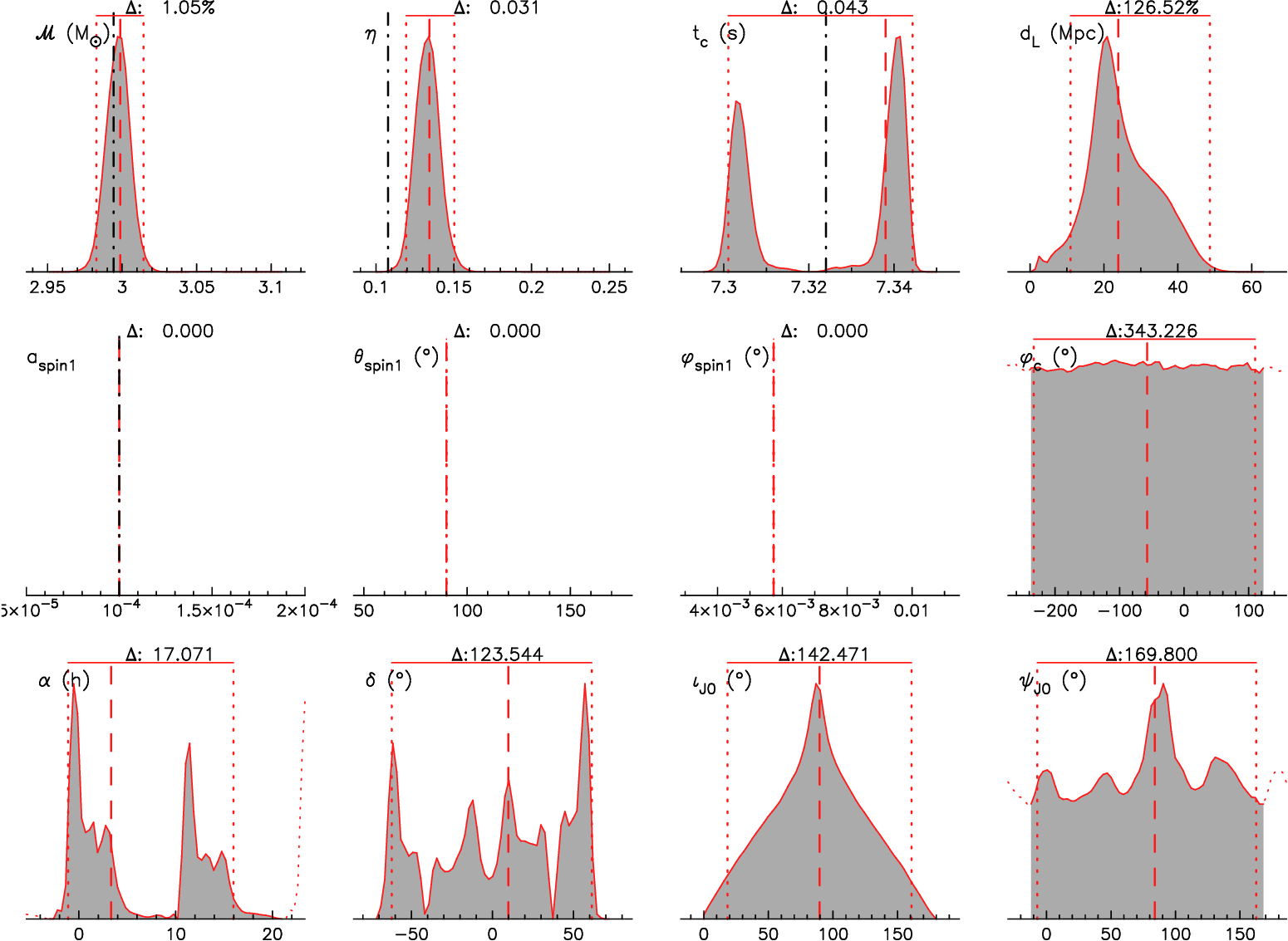}
  }
  \caption{
    One-dimensional PDFs showing the results of the nsMCMC analysis 
    of the BH-NS hardware injection.
    The meaning of the lines and numbers is the same as in 
    \fref{fig:bh-ns_pdf1d}.
  }
  \label{fig:bh-ns_pdf1d_ns}
\end{figure}

\Fref{fig:bh-ns_pdf1d_ns} contains the same information as 
\fref{fig:bh-ns_pdf1d}, but for the nsMCMC analysis.  Hence, the three
panels $a_\mathrm{spin,1}$, $\theta_\mathrm{spin,1}$ and $\varphi_\mathrm{spin,1}$
are empty. The medians and 2-$\sigma$ ranges for the main parameters in both the 
spMCMC and nsMCMC analyses are summarised in \tref{tab:bh-ns}.
We see clearly that the symmetric mass ratio $\eta$ is overestimated in both
parameter-estimation runs.  A discussion of a possible explanation of the biases
in this and the next analyses can be found in 
\sref{sec:conclusions}.  In addition to that, the PDF for $\eta$ is bimodal
for the spMCMC run, but not for the nsMCMC run.  This degeneracy seems to be
introduced by the correlations between $\eta$ and the spin 
parameters~\cite{2008ApJ...688L..61V}.  These correlations can account for
the larger uncertainties in the determination of $\mathcal{M}$ and $\eta$
in the spMCMC analysis.

\begin{table}
  \caption{
    Medians and 2-$\sigma$ ranges from the BH-NS analysis for selected parameters.
    \label{tab:bh-ns}
  }
  \begin{indented}
    \item[]
      \begin{tabular}{@{}lllllll}
        \br
                             & \multicolumn{2}{c}{$\mathbf{\mathcal{M}~(M_\odot)}$} & \multicolumn{2}{c}{$\mathbf{\eta}$}       & \multicolumn{2}{c}{$\mathbf{a_\mathrm{spin,1}}$}    \\
                             & med.    & 2-$\sigma$ range                           & med.    & 2-$\sigma$ range                & med.    & 2-$\sigma$ range   \\
        \mr
        \textbf{Injection}   & 2.994   & ---                                        & 0.107   & ---                             & 0.00    & ---                \\
        \textbf{spMCMC}      & 2.998   & 2.978 -- 3.016                             & 0.143   & 0.115 -- 0.248                  & 0.08    & 0.00 -- 0.94       \\
        \textbf{nsMCMC}      & 2.999   & 2.983 -- 3.014                             & 0.134   & 0.119 -- 0.150                  & ---     & ---                \\
        \br
      \end{tabular}
  \end{indented}
\end{table}

\begin{figure}
  \begin{center}
    \resizebox{\textwidth}{!}{
      \includegraphics[angle=0]{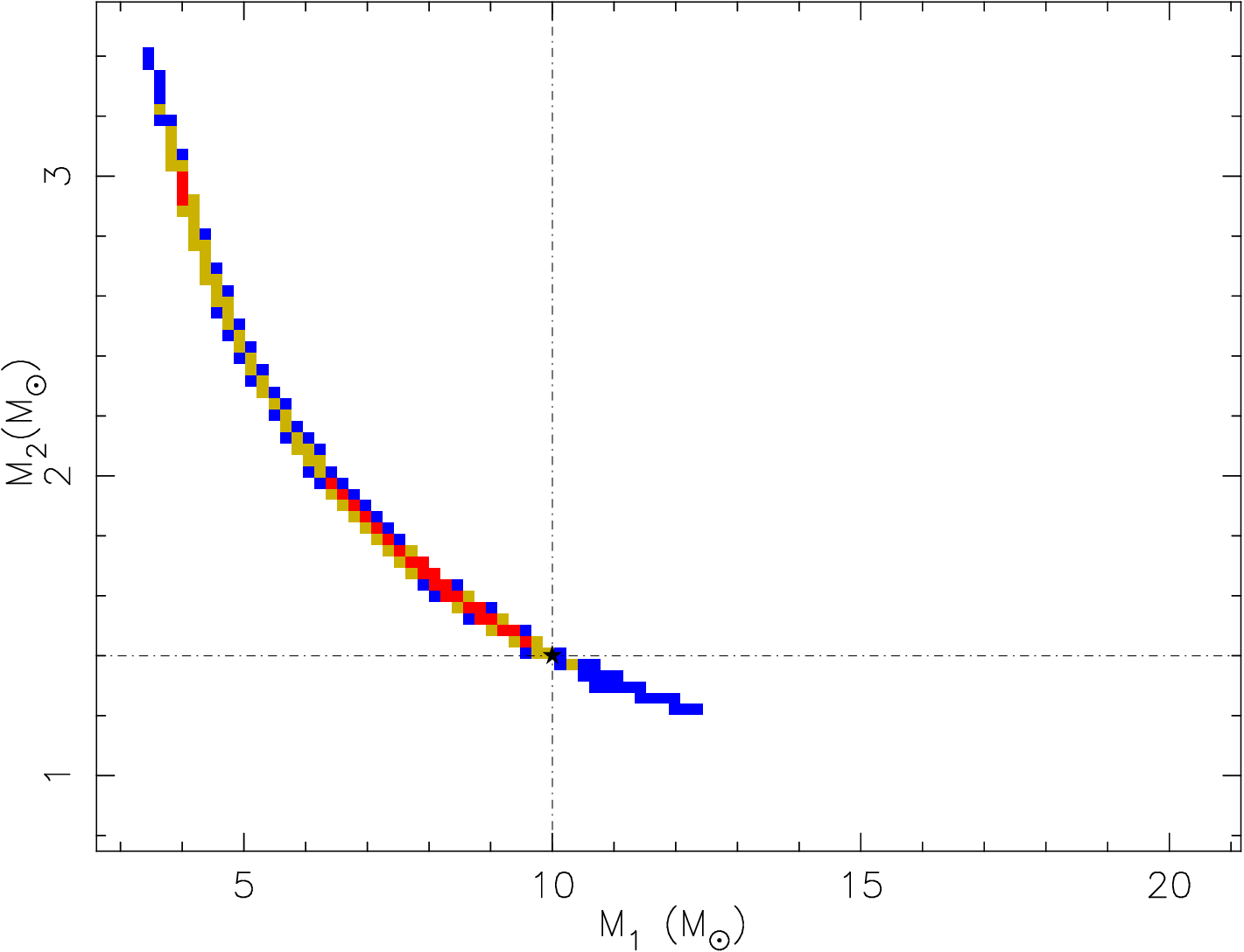}
      \hspace{0.1\textwidth}
      \includegraphics[angle=0]{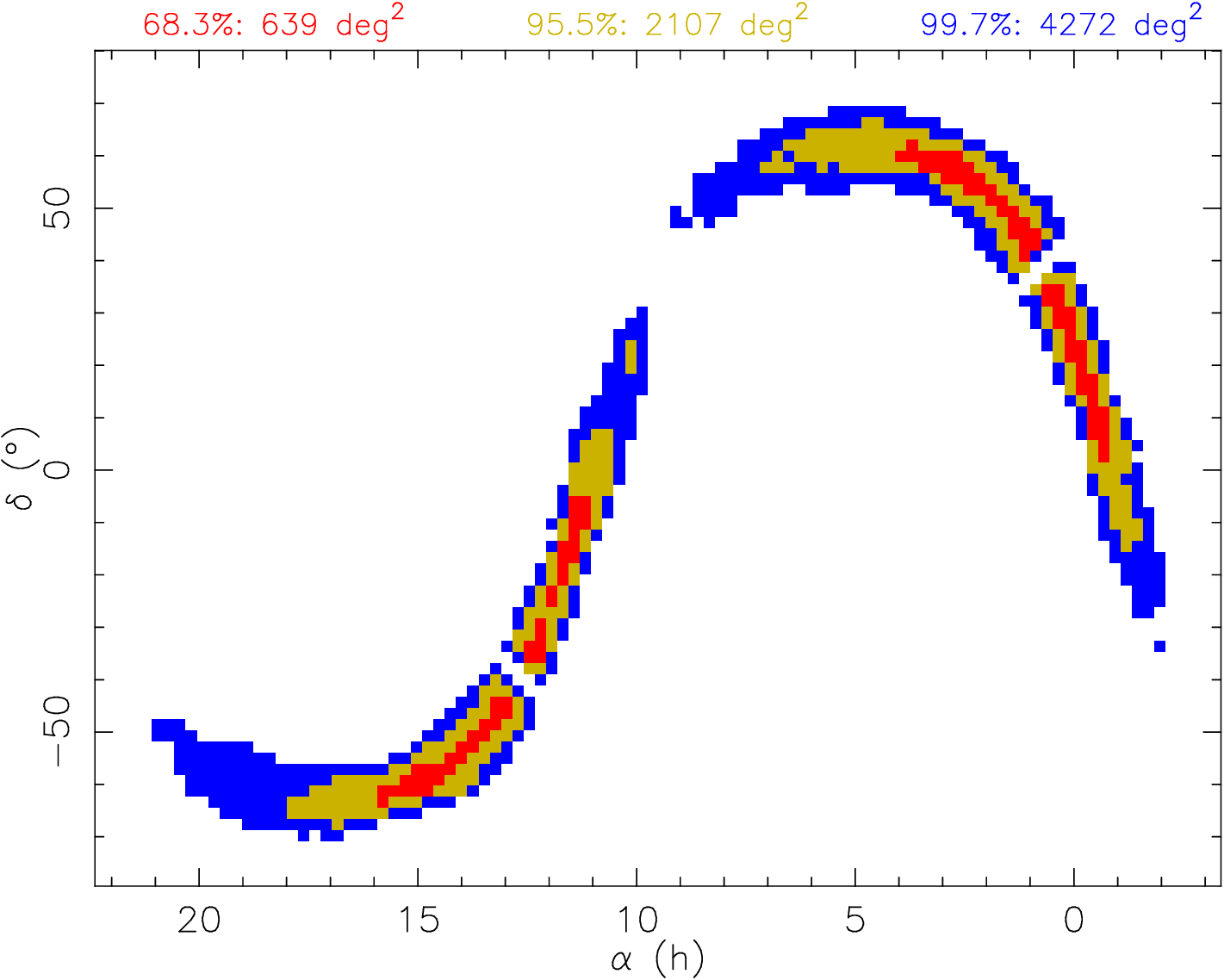}
    }
    \caption{
      Two-dimensional PDFs for the two individual masses (left panel, \textbf{(a)}) and 
      the position in the sky (right panel, \textbf{(b)}) in the spMCMC analysis of the 
      BH-NS inspiral.  The different colours/shades 
      show the 1-$\sigma$ (68.3\%), 2-$\sigma$ (95.4\%) and 3-$\sigma$ (99.7\%) 
      probability areas.
      The dashed lines in the left panel denote the masses of the injected 
      signal, the numbers in the header of the right panel indicate the surface 
      for each probability area in square degrees.\par}
    \label{fig:bh-ns_pdf2d}
  \end{center}
\end{figure}

\Fref{fig:bh-ns_pdf2d} shows two-dimensional PDFs from the spMCMC 
analysis of the BH-NS inspiral.   These PDFs are rather narrow and 
the surface contained in them is much smaller than suggested by the 
corresponding one-dimensional PDFs.  \Fref{fig:bh-ns_pdf2d}a shows a very 
thin arc in the $M_1$-$M_2$ plane, the width of which is determined by the uncertainty in
the chirp mass, and whose length is set by the much larger uncertainty
in $\eta$.  The 2D PDF in \fref{fig:bh-ns_pdf2d}b represents
a great circle in the sky, as expected for a coincident but non-coherent 
injection.  The gaps in the `ring' represent areas where the probability
density is low, due to a smaller volume of support in other extrinsic 
parameters, such as the binary orientation and distance~\cite{2008arXiv0812.4302R}.

\subsection{A neutron-star--neutron-star injection}
\label{sec:ns-ns}

We analysed the hardware injection that was performed
for the signal of the binary inspiral of two 1.4\,$M_\odot$ neutron stars 
(NSs) at a distance of 40\,Mpc, using 14.0\,s  of data from each of the 
detectors H1 and L1.
The results of both the spMCMC and nsMCMC analyses for this injection 
are summarised in \tref{tab:ns-ns}.  In this case of an equal-mass
injection, we cannot overestimate the value of $\eta$.  Instead, we find
that the chirp mass is underestimated in the nsMCMC run, but not in the
spMCMC run.  It is possible that an overestimation of $a_\mathrm{spin,1}$ 
in the spMCMC run mimics the effect of the underestimation of $\mathcal{M}$ 
in the nsMCMC run.  When compared to 
\tref{tab:bh-ns}, we see that the median for $a_\mathrm{spin,1}$
is significantly higher here, indicating that this parameter does not
reach the lower limit of the prior as much as is the case for the $a_\mathrm{spin,1}$
PDF of the BH-NS analyses in \fref{fig:bh-ns_pdf1d}.

\begin{table}
  \caption{
    Medians and 2-$\sigma$ ranges from the NS-NS analysis for selected parameters.
    \label{tab:ns-ns}
  }
  \begin{indented}
    \item[]
      \begin{tabular}{@{}lllllll}
        \br
                             & \multicolumn{2}{c}{$\mathbf{\mathcal{M} (M_\odot)}$}  & \multicolumn{2}{c}{$\mathbf{\eta}$}       & \multicolumn{2}{c}{$\mathbf{a_\mathrm{spin,1}}$}    \\
                             & med.    & 2-$\sigma$ range                            & med.    & 2-$\sigma$ range                & med.    & 2-$\sigma$ range   \\
        \mr
        \textbf{Injection}   & 1.219   & ---                                         & 0.250   & ---                             & 0.00    & ---                \\
        \textbf{spMCMC}      & 1.218   & 1.211 -- 1.229                              & 0.219   & 0.057 -- 0.250                  & 0.41    & 0.00 -- 0.80       \\
        \textbf{nsMCMC}      & 1.214   & 1.211 -- 1.216                              & 0.238   & 0.201 -- 0.250                  & ---     & ---                \\
        \br
      \end{tabular}
  \end{indented}
\end{table}

\subsection{A black-hole--black-hole injection}
\label{sec:bh-bh}

The LSC did a hardware injection into LIGO data for a 
3.0+3.0\,$M_\odot$, non-spinning BH-BH inspiral at a distance of 40\,Mpc.
We used 8.0\,s of data from 
both H1 and L1 for the analysis.
A summary of the results of the spMCMC and nsMCMC analyses is listed in \tref{tab:bh-bh}.
As for the NS-NS result in \sref{sec:ns-ns}, we find that the chirp 
mass is underestimated for the nsMCMC run, but not for the spMCMC run.
In the spMCMC run, the value for $a_\mathrm{spin,1}$ is again higher than
the recovered value in the analysis of the BH-NS signal, and here this
parameter is actually slightly overestimated.  In \sref{sec:conclusions}
we discuss a possible explanation for these biases.

\begin{table}
  \caption{
    Medians and 2-$\sigma$ ranges from the BH-BH analysis for selected parameters.
    \label{tab:bh-bh}
  }
  \begin{indented}
    \item[]
      \begin{tabular}{@{}lllllll}
        \br
                             & \multicolumn{2}{c}{$\mathbf{\mathcal{M} (M_\odot)}$}  & \multicolumn{2}{c}{$\mathbf{\eta}$}       & \multicolumn{2}{c}{$\mathbf{a_\mathrm{spin,1}}$}    \\
                             & med.    & 2-$\sigma$ range                            & med.    & 2-$\sigma$ range                & med.    & 2-$\sigma$ range   \\
        \mr
        \textbf{Injection}   & 2.612   & ---                                         & 0.250   & ---                             & 0.00    & ---                \\
        \textbf{spMCMC}      & 2.608   & 2.576 -- 2.634                              & 0.236   & 0.153 -- 0.250                  & 0.35    & 0.03 -- 0.73       \\
        \textbf{nsMCMC}      & 2.588   & 2.576 -- 2.595                              & 0.240   & 0.212 -- 0.250                  & ---     & ---                \\
        \br
      \end{tabular}
  \end{indented}
\end{table}

\section{Conclusions and future work}
\label{sec:conclusions}

In these proceedings, we showed that our MCMC code \textsc{SPINspiral} can provide parameter
estimation for binary-inspiral signals under realistic circumstances.
Whereas in \cite{2008ApJ...688L..61V} we tested the parameter-estimation
code on software injections that we performed ourselves in Gaussian
noise and with a fairly high SNR (17 for the network), the examples 
in this paper show that our code is capable of the post-detection 
follow-up analysis it was designed for.  The hardware injections were
performed with a different waveform family, with relatively low SNRs 
and in observed LIGO detector data.  We demonstrated that our parameter-estimation code can use the
information that could be available from a detection trigger and return
the posterior-density functions for the physical parameters.  The 
results for the weaker signals in our analysis suggest that 
\textsc{SPINspiral} is able to analyse even the weakest signals that will be detected
by the CBC detection pipeline.

Because the hardware injections were done non-coherently, the results presented 
in \sref{sec:analysis} are of limited interest for most parameters.  
The posterior PDFs for the sky location should be a great circle, which is what we find.
However, the inclination $\iota_\mathrm{J_0}$ should indicate a `face-on' system, 
whereas our PDFs prefer an `edge-on' system, either because of the shape of the prior 
distribution, or more strongly than that.
The recovered PDFs for the phases $\varphi_\mathrm{spin,1}$ and $\varphi_\mathrm{c}$, 
and the polarisation angle $\psi_\mathrm{J_0}$ are practically meaningless
here.  Because the hardware injections were overhead in each 
detector, there is no consistent solution for $t_\mathrm{c}$, and the 
recovered distance is smaller than the distance of the injection.

The results for the mass and spin parameters are much more interesting.
In several cases, our PDFs for some of these parameters are offset from the injection 
values. The main reason for this is most likely the systematic error 
that is introduced by the difference in waveforms used for the injection (2.0-pN)
and the parameter estimation (1.5-pN).  This difference may give rise to a
difference of $\sim 10$ gravitational-wave cycles during the inspiral in the LIGO band~\cite{lrr-2006-4}. 
It seems reasonable that parameters like $\mathcal{M}$ and $\eta$ should be
changed significantly in order to compensate for this effect.  

In the BH-NS inspiral in \sref{sec:bh-ns} we clearly overestimate $\eta$ 
in both the spMCMC and nsMCMC runs.  In the other two analyses, $\eta$ cannot
be overestimated, and we find biases in $\mathcal{M}$ or $a_\mathrm{spin,1}$
instead, which may be due to the correlations between these parameters.
On the other hand, we expect that the statistical accuracies which we find in the parameter 
estimation are representative, and that these accuracies will improve once the 
systematic biases are dealt with.

To achieve this improvement, part of our current work focuses on the im\-ple\-men\-ta\-tion of a more 
realistic waveform template. \textsc{SPINspiral} is now able to use a 3.5-pN template 
from the LSC Algorithm Library~\cite{lal}, which allows for the spins of both 
binary members \cite{2003PhRvD..67j4025B}.  Our first results indeed show an 
improvement in accuracy, especially in the mass parameters, due to the removal of the
systematic errors mentioned here (using the same waveform family for injections and 
parameter-estimation templates),
and due to the higher post-Newtonian order used~\cite{2008arXiv0812.4302R}.

Our second focus of code development is aimed at adding \textsc{SPINspiral} to 
the CBC follow-up pipeline.  This will allow many people to run the code ---
and to run it automatically --- on triggers that are plausible 
inspiral events (either from gravitational waves, or from hardware injections).  
Inclusion of \textsc{SPINspiral} in the pipeline will increase the number of tests we 
can perform, and make these tests as realistic as possible, optimally
approximating the `real' process of gravitational-wave detection and analysis.

\section*{Acknowledgements}

This work is partially supported by a Packard Foundation Fellowship 
and a NSF Gravitational Physics grant 
(PHY-0653321) to VK; NSF Gravitational  Physics grant PHY-0553422 to NC. 
Computations were performed on the Fugu computer cluster funded by NSF MRI 
grant PHY-0619274 to VK.

\section*{References}
\bibliographystyle{unsrt}  
\bibliography{gwdaw13}

\end{document}